\documentstyle[aps,prb,epsfig]{revtex}
\begin{document}
\baselineskip=0.5cm
\renewcommand{\thefigure}{\arabic{figure}}
\title{Pair densities at contact in the quantum electron gas}
\author{R. Asgari$^{1,2}$, M. Polini$^1$, B. Davoudi$^{1,2}$, and M. P. Tosi$^1$~\footnote{Corresponding author: tosim@sns.it}}
\address{
$^1$NEST-INFM and Classe di Scienze, Scuola Normale Superiore, I-56126 Pisa, Italy\\ 
$^2$Institute for Studies in Theoretical Physics and Mathematics, Tehran, P.O.Box 19395-5531,Iran\\
} 
\maketitle
\vspace{1 cm}

{\bf Abstract}
\vspace{0.2 cm}

The value of the pair distribution function $g(r)$ at contact $(r = 0)$ in a quantum electron gas 
is determined by the scattering events between pairs of electrons with antiparallel spins. The 
theoretical results for $g(0)$ as a function of the coupling strength $r_s$ in the paramagnetic electron 
gas in dimensionality $D=2$ and $3$, that have been obtained from the solution of the two-body 
scattering problem with a variety of effective scattering potentials embodying many-body effects, 
are compared with the results of many-body calculations in the ladder approximation and with 
quantum Monte Carlo data.
\vspace{0.3cm}


\newpage

The equilibrium pair distribution function $g(r)$ in a quantum electron gas describes the 
so-called Pauli and Coulomb hole which on average surrounds each electron from many-body 
exchange and Coulomb correlations. This function, which has an important role in the 
development of non-local density functional theories~[1], is a weighted mean of the functions $g_{\uparrow\uparrow}(r)$  and $g_{\uparrow\downarrow}(r)$ 
for pairs of electrons with parallel or antiparallel spins. On account of the 
Pauli principle, $g_{\uparrow\uparrow}(r)$ vanishes at contact $(r=0)$ while $g_{\uparrow\downarrow}(0)$ 
is wholly determined by scattering events between electron pairs. 
The value of $g_{\uparrow\downarrow}(0)$ enters to determine the 
asymptotic values of the local field factors at large momenta and thus has a specific role in 
describing exchange and short-range correlation effects in the linear response properties of the 
electron gas (see {\it e.g.}~[2] and references given therein).

Although the value of $g_{\uparrow\downarrow}(0)$ results strictly from two-body collisional events, as noted above, the many-body aspects of the interactions between electrons in the quantum Coulomb gas 
affect the pair wave functions in a profound way. Two main theoretical approaches have been 
followed in the literature for evaluating $g_{\uparrow\downarrow}(0)$. 
The first is based on many-body diagrammatic 
techniques and the second involves the direct calculation of the pair wave function from an 
appropriate Schr\"odinger equation. The main purpose of this Letter is to compare the results that 
have been obtained by these two methods in the paramagnetic fluid state over a broad range of 
values of the coupling strength $r_s$ and to assess their validity against the quantum Monte Carlo 
(QMC) data that have become available. The main emphasis will be on results for the electron 
gas in dimensionality $D=2$, where the many-body effects at any given $r_s$ are comparatively 
more important. A discussion will also be given, however, for the 3D case.

A brief presentation of the two theoretical approaches will be useful. The approach first 
proposed by Yasuhara~[3] and by Hede and Carbotte~[4] for the 3D electron gas evaluates the 
ladder diagrams representing the virtual processes in which two electrons are excited from the 
unperturbed Fermi sphere leaving two holes behind and repeat their mutual interactions {\it via} the 
Coulomb potential to finally be scattered back into their initial momentum states. The calculation 
of these diagrams leads to an effective interaction between the two electrons which is to be 
determined from the solution of a Bethe-Goldstone integral equation. A simple analytic 
expression is obtained for $g_{\uparrow\downarrow}(0)$ in terms of a modified Bessel function when the kernel of the integral equation is approximated to include the screening effects in a somewhat crude way. 
Further work on the electron-electron interactions in $D=3$ from the Bethe-Goldstone equation 
was carried out by Lowy and Brown~[5], who were able to establish a close connection between 
the diagrammatic approach and the self-consistent method of Singwi {\it et al.}~[6].

The alternative method for the evaluation of pair distribution functions in the electron gas 
is based on the solution of the Schr\"odinger equation for the electron-pair wave function with the 
use of an effective scattering potential embodying many-body effects. This approach was 
initiated in the work of Overhauser~[7], who constructed a simple model for the scattering 
potential in 3D as the electrostatic potential of an electron surrounded by a Wigner-Seitz sphere 
of neutralizing background. This potential vanishes outside the sphere, allowing simple analytic 
expressions to be obtained in this way for $g_{\uparrow\downarrow}(0)$ and for the $s$-wave scattering length~[7]. Later refinements have involved an accurate numerical solution of Overhauser's two-body Schr\"odinger equation~[8] and a self-consistent Hartree model for the scattering potential~[9].

In the pair-scattering approach the value of $g_{\uparrow\downarrow}(0)$ is determined by the square modulus of the $s$-wave component of the pair wave function $\Phi^{(\uparrow\downarrow)}_{k, \ell}(r)$, 
averaged over the probability $p(k)$ of finding two electrons with given relative momentum $k$: that is,
\begin{equation}\label{e1}
g_{\uparrow\downarrow}(0)=\frac{1}{r}\langle|\Phi^{(\uparrow\downarrow)}_{k, \ell=0}(0)|^2 \rangle_{p(k)}
\end{equation}
in $D=2$. Here, $\Phi^{\uparrow\downarrow}_{k, \ell=0}(r)$ is the solution of the Schr\"odinger equation
\begin{equation}\label{e2}
\left\{-\frac{\hbar^2}{m} \frac{d^2}{d r^2} -\frac{\hbar^2}{4m\,r^2}+V^{(\uparrow\downarrow)}_{\rm \scriptstyle KS}(r)\right\}\Phi^{(\uparrow\downarrow)}_{k, \ell=0}(r)=\frac{\hbar^2k^2}{m}\,\Phi^{(\uparrow\downarrow)}_{k, \ell=0}(r) 
\end{equation}	
where $V^{(\uparrow\downarrow)}_{\rm \scriptstyle KS}(r)$ has been shown to be the Kohn-Sham effective potential associated with the local inhomogeneity in the density of electrons with given spin surrounding an electron of opposite spin~[10]. The function $p(k)$ can be evaluated from the free-electron momentum 
distribution, the result obtained by Ziesche {\it et al.}~[11] for the 2D electron gas being
\begin{equation}\label{e3}
p(k)=\frac{16\,k}{\pi\,k^2_F}\,\left[\arccos{\left(\frac{k}{k_{F}}\right)}-\frac{k}{k_F}\,\sqrt{1-\frac{k^2}{k^2_F}}~\right]~.
\end{equation}		              
Here, $k_F=\sqrt{2}/(r_s a_B)$ is the Fermi wave number, with $r_s a_B$ related to the areal density $n$ of 
electrons by $r_s a_B=(\pi n)^{-1/2}$. The determination of $g_{\uparrow\downarrow}(0)$  
thus hinges in this method on a suitable approximation for the scattering potential $V^{(\uparrow\downarrow)}_{\rm \scriptstyle KS}(r)$.	

A direct extension of Overhauser's model to the 2D case suffers from the difficulty that 
the electrical potential due to an electron plus its neutralizing Wigner-Seitz disc does not vanish 
outside the disc~[12, 13]. Furthermore, inclusion of exchange and correlation through a spin-dependent 
scattering potential is needed to account for the emergence of a 
first-neighbour shell already at relatively low values of the coupling strength in this case~[14]. We 
report in Figure 1 the results that we have obtained from Eq.~(2) for $g_{\uparrow\downarrow}(0)$ 
in 2D, by using the self-consistent spin-dependent potentials proposed in Ref.~[10]. 
That is,
\begin{equation}\label{e4}
V^{(\sigma\sigma')}_{\scriptstyle \rm KS}(q)=v(q)+\sum_{\sigma''}v(q)[1-G_{\sigma\sigma''}(q)]\,[S_{\sigma''\sigma'}(q)-\delta_{\sigma''\sigma'}]
\end{equation}
where the sum runs over the two spin orientations, $v(q)=2 \pi e^2/q$ is the bare Coulomb 
potential, and $G_{\sigma\sigma'}$ are the local-field factors describing the effects of exchange 
and short-range correlations in the dielectric and magnetic response of the 2D electron gas, that we have 
taken from QMC data of Moroni {\it et al.}~[15, 16] as fitted by interpolation formulae in Reff.~[17,18]. Finally, the functions $S_{\sigma\sigma'}(q)$ in Eq.~(4) are the partial liquid structure factors, which are 
related to the spin-resolved pair distribution functions by
\begin{equation}
S_{\sigma\sigma'}(q)=\delta_{\sigma\sigma'}+\frac{n}{2}\int d^{2} {\bf r}\,[g_{\sigma\sigma'}(r)-1]\exp{(-i {\bf q} \cdot {\bf r})}~.
\end{equation}
The dependence of the scattering potentials in Eq.~(4) on the pair distribution functions requires 
a self-consistent solution of the effective two-body Schr\"odinger equations for both  $g_{\uparrow\uparrow}(r)$ and $g_{\uparrow\downarrow}(r)$ (see Ref.~[10]).

Figure~1 plots the quantity $r_sg(0)=r_sg_{\uparrow\downarrow}(0)/2$ as a function of the coupling strength $r_s$, as a device for emphasizing the large-$r_s$ region where $g_{\uparrow\downarrow}(0)$ 
is becoming very small at the expense of the low-$r_s$ region where $g_{\uparrow\downarrow}(0)$ can be calculated perturbatively~[12]. Together with our present results we report those calculated by Isawa and Yasuhara~[19] and by Nagano {\it et al.}~[20,21] 
in the ladder approximation and those obtained in a QMC study by Dr. S. Moroni (private 
communication). The agreement between these two basic theoretical approaches with each other 
and with the QMC data is quite remarkable. It is also worth noticing from Figure 1 
that the use of the Hartree 
model in the scattering approach~[9] leads to results of good quantitative value for $g_{\uparrow\downarrow}(0)$.
 
Figure~1 reports also some other theoretical results, which appear to be rather different 
from the evidence discussed just above. These are (i) the results reported by Freeman~[22] from 
a coupled-cluster summation of ladder diagrams; (ii) the results obtained by representing the 
scattering potential in 2D through simple electrostatic models~[12, 13]; and (iii) the results 
taken up to $r_s \simeq 5$ from the work of Bulutay and Tanatar~[23], who used a classical-map hypernetted chain (CHNC) model 
proposed by Dharma-wardana and Perrot~[24].

A similar examination of theoretical values for $r_sg_{\uparrow\downarrow}(0)$ 
as a function of $r_s$  in the 3D electron gas is displayed in Figure~2. 
While in this case the lack of data on local-field factors has 
prevented a self-consistent calculation including exchange and correlation in the scattering 
potentials, Figure 2 shows that the Hartree model results~[9] are again in quite good agreement 
with those obtained by the evaluation of ladder diagrams~[3] and from the QMC study of Ortiz 
and Ballone~[25]. Figure 2 also reports the results obtained with Overhauser's electrostatic 
model for the scattering potential~[7, 8], which in the 3D case are in good agreement with the 
other evidence discussed just above. There are appreciable discrepancies, however, with the 
predictions given by an interpolation formula reported by Gori-Giorgi {\it et al.}~[26] from 
unpublished QMC work by Ortiz and coworkers.

In summary, we have discussed the status of various theoretical approaches to the 
calculation of the pair distribution function at contact in the electron gas. In view of the 
demonstrated sensitivity of this quantity to the details of the theory and to the role of exchange 
and short-range correlations, especially in the case $D=2$, further accurate studies by quantum simulation 
techniques would be useful.
\vspace{0.5 cm}

\noindent {\bf Ackowledgements}

This work was partially funded by MIUR under the PRIN2001 Initiative. We thank Dr. S. Moroni for providing us with his unpublished QMC data reported in Figure 1.
\newpage

{\bf References}
\vspace{0.2 cm}

[1] O. Gunnarsson, M. Jonson, B.I. Lundqvist, Phys. Rev. B 20 (1979) 3136.

[2] M. Polini, M.P. Tosi, Phys. Rev. B 63 (2001) 045118.

[3] H. Yasuhara, Solid State Commun. 11 (1972) 1481.

[4]	B.B.J. Hede, J.P. Carbotte, Can. J. Phys. 50 (1972) 1756.

[5]	D.N. Lowy, G.E. Brown, Phys. Rev. B 12 (1975) 2138.

[6]	K.S. Singwi, M.P. Tosi, R.H. Land, A. Sj\"olander, Phys. Rev. 176 (1968) 589.

[7]	A.W. Overhauser, Can. J. Phys. 73 (1995) 683.

[8]	P. Gori-Giorgi, J.P. Perdew, Phys. Rev. B 64 (2001) 155102.

[9]	B. Davoudi, M. Polini, R. Asgari, M.P. Tosi, Phys. Rev. B 66 (2002) 075110.

[10]	B. Davoudi, M. Polini, R. Asgari, M.P. Tosi, cond-mat/0206456.

[11]	P. Ziesche, J. Tao, M. Seidl, J.P. Perdew, Int. J. Quantum Chem. 77 (2000) 819.

[12]	M. Polini, G. Sica, B. Davoudi, M.P. Tosi, J. Phys.: Condens. Matter 13 (2001) 3591.

[13]	J. Moreno, D.C. Marinescu, cond-mat/0210137.

[14]	F. Capurro, R. Asgari, B. Davoudi, M. Polini, M.P. Tosi, Z. Naturforsch. 57a (2002) 237.

[15]	S. Moroni, D.M. Ceperley, G. Senatore, Phys. Rev. Lett. 69 (1992) 1837.

[16]	S. Moroni, D.M. Ceperley, G. Senatore, Phys. Rev. Lett. 75 (1995) 689.

[17]	B. Davoudi, M. Polini, G.F. Giuliani, M.P. Tosi, Phys. Rev. B 64 (2001) 153101.

[18]	B. Davoudi, M. Polini, G.F. Giuliani, M.P. Tosi, Phys. Rev. B 64 (2001) 233110.

[19]	Y. Isawa, H. Yasuhara, Solid State Commun. 46 (1983) 807.

[20]	S. Nagano, K.S. Singwi, S. Ohnishi, Phys. Rev. B 29 (1984) 1209.

[21]	S. Nagano, K.S. Singwi, S. Ohnishi, Phys. Rev. B 31 (1985) 3166.

[22]	D.L. Freeman, J. Phys. C 16 (1983) 711.

[23]	C. Bulutay, B. Tanatar, Phys. Rev. B 65 (2002) 195116.

[24]	M.W.C. Dharma-wardana, F. Perrot, Phys. Rev. Lett. 84 (2000) 959.

[25]	G. Ortiz, P. Ballone, Phys. Rev. B 50 (1994) 1391.

[26]	P. Gori-Giorgi, F. Sacchetti, G.B. Bachelet, Phys. Rev. B 61 (2000) 7353.
\newpage

\begin{figure}
\centerline{\mbox{\psfig{figure=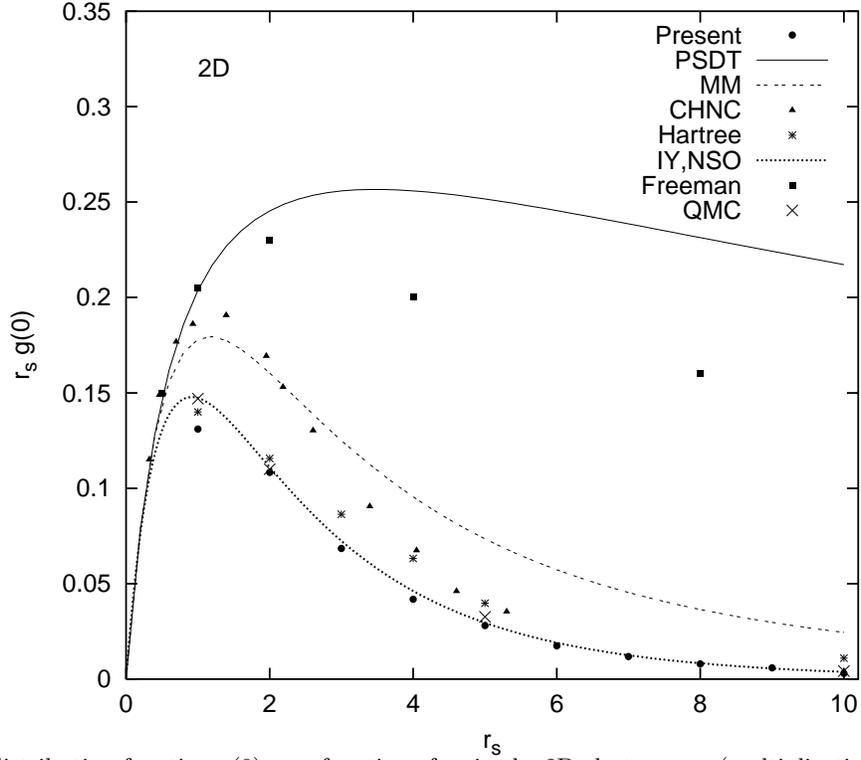, angle =0, width =12 cm}}} 
\caption{The pair distribution function $g(0)$ as a function of $r_s$  in the 2D electron gas (multiplication of $g(0)$ by $r_s$ enhances the strong-coupling regime). The 
present results are shown as dots and the QMC data as crosses. The meaning of the other 
symbols is as follows: PSDT and MM from Reff.~[12] and~[13], respectively; CHNC, from 
Ref.~[23]; Hartree, from Ref.~[9]; IY, NSO from Reff.~[19-21]; Freeman, from Ref.~[22]. 
The present results for $r_s g(0)$, as well as the QMC and Hartree ones, involve an extrapolation of 
the values of $g_{\uparrow\downarrow}(0)$ to $r=0$, that we have carried out with the help of the cusp-condition law, $g_{\uparrow\downarrow}(r)=g_{\uparrow\downarrow}(0)[1+\sqrt{2}~r_s~(r k_F)+...]$.}
\label{Fig1}
\end{figure}
\newpage

\begin{figure}
\centerline{\mbox{\psfig{figure=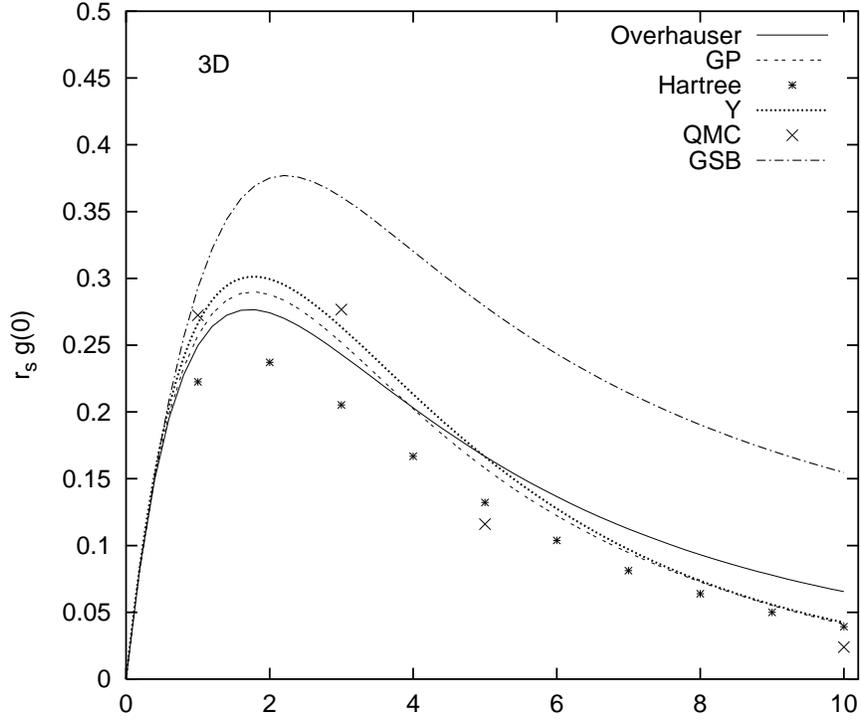, angle =0, width =12 cm}}} 
\caption{The pair distribution function $r_s g(0)$ as a function of $r_s$  in the 3D electron gas. 
The QMC data from Ref.~[25] are shown as crosses. The meaning of the other symbols is as 
follows: Overhauser and GP from Reff.~[7] and~[8], respectively; Hartree, from Ref.~[9]; Y from 
Ref.~[3]; GSB, from Ref.~[26]. The cusp-condition law in 3D is $g_{\uparrow\downarrow}(r)=g_{\uparrow\downarrow}(0)[1+(9\pi/4)^{-1/3}~r_s~(r k_F)+...]$.}
\label{Fig2}
\end{figure}
\end{document}